\documentclass{jaa}
\usepackage{natbib}
\usepackage{hyperref}
\usepackage{longtable}
\usepackage{natbib}
\usepackage{booktabs}

\usepackage{graphicx}

\begin{document}\sloppy

\title{DHARA: Data Handling and Automated Reduction pipeline for AIMPOL}



\author{Namita Uppal\textsuperscript{1,2,*}, Shashikiran Ganesh\textsuperscript{2}, Santosh Joshi\textsuperscript{3}, Dmitry Blinov\textsuperscript{1}, Konstantinos Tassis\textsuperscript{1,4}, \and Sadhana Singh\textsuperscript{5}}
\affilOne{\textsuperscript{1} Institute of Astrophysics, Foundation for Research and Technology - Hellas, Vasilika Vouton, GR-70013 Heraklion, Greece\\}
\affilTwo{\textsuperscript{2}Physical Research Laboratory, Ahmedabad, 380009, Gujarat, India\\}
\affilThree{\textsuperscript{3}Aryabhatta Research Institute of Observational Sciences, Manora Peak, Nainital, Uttarakhand-263001, India\\}
\affilFour{\textsuperscript{4}Department of Physics and Institute of Theoretical and Computational Physics, University of Crete, Voutes Campus, 70013 Heraklion, Greece \\}
\affilFive{\textsuperscript{5}Department of Physics, Maulana Azad National Institute of Technology  , Bhopal-462003, India}


\twocolumn[{

\maketitle

\corres{namita@ia.forth.gr}


\begin{abstract}
We present an automated data-reduction and analysis  pipeline\footnote{The pipeline source code can be accessed using the GitHub link: \href{https://github.com/NamitaU/DHARA}{https://github.com/NamitaU/DHARA}. An archived version corresponding to this study is available at Zenodo: \href{https://doi.org/10.5281/zenodo.21158246}{https://doi.org/10.5281/zenodo.21158246}} for optical linear polarimetric data obtained from a dual-beam polarimeter. The pipeline is optimized for observations acquired with the ARIES Imaging POLarimeter mounted at the Cassegrain focus of the 1.04-m Sampurnanand Telescope at ARIES.
It is implemented using interactive Python routines that
process raw images to derive the Stokes parameters, from which the degree of polarization and polarization angle are computed along with their uncertainties. 
The pipeline framework is designed to handle the reduction of both single-source and crowded-field observations. The identification of extraordinary (e-ray) and ordinary (o-ray) image pairs is a crucial step and is performed using different strategies for single-source and crowded-field data, while subsequent stages, such as photometric and polarimetric analysis, follow a common procedure.
We validate the pipeline using observations of polarized standard stars acquired over multiple epochs between 2017 to 2025, and compare the results with values reported in the literature. To demonstrate the applicability of the pipeline to crowded-field observations, we apply it to polarimetric data of the Alessi~1 open cluster and compare with previously published results derived using traditional reduction methods.
In both cases, the polarization parameters derived using the pipeline agree with literature values within  $2\sigma$ uncertainties. Although developed for AIMPOL, the pipeline is readily adaptable to any dual-channel imaging polarimeter in which the e-ray and o-ray images are recorded in a FITS image from a single CCD. 
 
\end{abstract}


\keywords{techniques: polarimetric---methods:
observational---instrumentation: polarimeters.}
}]


\doinum{12.3456/s78910-011-012-3}
\artcitid{\#\#\#\#}
\volnum{000}
\year{0000}
\pgrange{1--}
\setcounter{page}{1}
\lp{1}

\section{Introduction}\label{sec:1}
The foundation of observational astrophysics largely relies on measuring the radiation flux or brightness of celestial objects, either through broadband photometry or spectroscopy. However, light from astronomical sources is often partially polarized. Polarimetry provides an additional diagnostic dimension beyond traditional photometric and spectroscopic observations, offering insights into the physical properties, geometry, and magnetic properties of the emitting or intervening medium. The polarization state of light encodes information that enables detailed investigations of various astrophysical environments, such as the interstellar medium (ISM), the cosmic microwave background, active galactic nuclei,  magnetic cataclysmic variables or other variable stars and solar system objects \citep[e.g.,][and reference therein]{Tassis2015, Joshi2017, Uppal2022a, Kochkina2023, Liodakis2023, Laura2023, Versteeg2023, Blinov2024, Uppal2024, Doi2024,  Dinsmore2025, Maxim2025, Tahani2025}. 

In the past few decades, many studies have been conducted to measure the state of polarization of the astronomical objects using imaging polarimetry or spectropolarimetry \citep[e.g.,][]{Pelgrims2024, Yenifer2025, pelgrims2025, Cikota2025, Thomson2025, Lieb2025}. In particular, polarization studies of the interstellar medium (ISM), focusing on dust grain physics \citep[][]{Andersson2015, Sadhana2022, Pravash2025} and the role of magnetic fields in star formation \citep[][]{Joshi1985, Soam2017, Choudhury2022, Rai2023, Paul2024, Barman2025, Gupta2025} through starlight polarization measurements have attracted significant attention within the Indian astronomical community. Starlight polarization was first discovered independently by \citet{Hall1949} and \citet{Hiltner1949}, and is understood to arise from non-spherical dust grains whose short axes tend to align parallel to the ambient magnetic field. As background starlight passes through such aligned grains, it becomes partially linearly polarized due to dichroic extinction \citep{Andersson2015}. Despite the long-standing interest in stellar polarimetry, most studies to date have been limited to star-forming regions with sparse sky coverage.

A few large-scale surveys have been initiated to address this gap. For example, a near-infrared (NIR) starlight polarization survey has been conducted across the Galactic disk \citep[GPIPS,][]{GPIPS, GPIPSDR4} between longitudes $18^\circ$ and $56^\circ$, while the ongoing SouthPOL survey \citep{Magalhaes2005} provides optical linear polarization data for the southern hemisphere. The forthcoming Polar-Areas Stellar-Imaging in Polarization High-Accuracy Experiment (PASIPHAE) survey \citep{Tassis2018} aims to cover high-latitude regions of the Galaxy in both hemispheres. However, no such survey is planned for the northern Galactic disk, leaving this region relatively unexplored.

Indian observational facilities equipped with polarimetric instruments are therefore expected to make important contributions by extending polarimetric coverage of the northern Galactic disk. In this context, earlier efforts in India include the dual-channel imaging polarimeter IMPOL, developed in the late 1990s and used extensively for various polarimetric observing campaigns \citep{Ram1998, Sen2000, sen2009, Goswami2010, Goswami2013}. Currently, only two instruments in India are actively measuring the linear polarization of starlight, one is EMCCD-based polarimeter \citep[EMPOL,][]{Ganesh2020} mounted on the 1.2-m telescope at  Mount Abu Observatory hosted by Physical Research Laboratory, and the other is ARIES Imaging Polarimeter \citep[AIMPOL,][]{Rautela2004, Pandey2023} installed at the 1.04-m Sampurnanand Telescope operated by ARIES. EMPOL is a single-channel polarimeter with a field of view (FOV) of approximately 3 arcmin, whereas AIMPOL is a dual-channel polarimeter offering a wider effective FOV of about 8 arcmin in diameter. Dual-beam imaging polarimeters are particularly advantageous for stellar polarimetric observations because they simultaneously record orthogonally polarized beams over an extended field, thereby improving observing efficiency and reducing the effects of atmospheric transparency variations.

Despite the advantages, dual-channel imaging polarimeters are limited not only in India but also worldwide, and are mostly available on small - and intermediate-aperture telescopes. Notable examples include IAGPOL \citep[Instituto de Astronomia, Geof\'isica e Ci\^encias Atmosf\'ericas (IAG) of the University of S\~ao Paulo (USP) POLarimeter,][]{iagpol} used on 0.9~m, 1.5~m, 1.6~m, and 0.6~m telescopes at CTIO (Cerro Tololo InterAmerican Observatory), ODP/LNA (Pico dos Dias Observatory Laborat\'orio Nacional de Astrof\'isica), and IAG observatory; Double imaging Polarimeter \citep[DIPOL,][]{dipol} on 60~cm KVA (Kungliga Vetenskapsakademien) telescope in Roque de los Muchachos (ORM) and its successors DIPOL-2 \citep{Dipol2} and DIPOL- Ultra Fast \citep[DIPOL-UF,][]{DipolUF} on various telescopes like 2.6~m NOT (Nordic Optical Telescope), 60~cm KVM telescope, 4.2~m William Herschel Telescope (WHT) at ORM observatory as well as 2.2~m UH88 telescope at Mounakea, and 1.27~m (H127) telescope in the Univerity of Tasmania; FOcal Reducer and low dispersion Spectrograph (FORS)~1 and 2 \citep{fors1, fors_manual_2002, Patat2006} on VLT, at Paranal observatory, Chile; Alhambra Faint Object Spectrograph and Camera \citep[ALFOSC,][]{alfosc_manual} on 2.6~m NOT telescope at ORM; and Hatfield polarimeter \citep[Hatpol,][]{Hough1991} on 3.8~m UKIRT (United Kingdom Infrared Telescope) telescope at Maunakea.

AIMPOL, represents an important dual-channel imaging polarimetric facility for optical stellar polarization studies in India.
The instrument has been extensively used for stellar polarimetric observations over the past two decades, e.g., \citep{Eswaraiah2012, Eswaraiah2013, Pandey2013, Soam2017, Ekta2020, Piyali2022,  Sadhana2020, Uppal2024, Neha2024, Bijas2024, Biswas2026}. However, the data reduction process has largely relied on manual or semi-automated procedures, that slows down
the science production. Such approaches are time-consuming and may introduce user-dependent variations in the derived polarization parameters. Moreover, with the increasing number of observations and the growing need for uniform and reproducible results, a standardized automatic data reduction framework has become essential. 

In recent years, several automated polarimetric reduction pipelines have been developed for different instruments. For example, SOLVEPOL \citep{solvepol} is an Interactive Data Language (IDL)-based automated reduction pipeline optimized for the IAGPOL instrument, while its Python-based implementation and expanded version are available through ASTROnomical POlarimetry and Photometry pipeline \citep[ASTROPOP,][]{astropop}. Similarly, PyNOT provides automated reduction tools for the ALFOSC instrument, and dedicated reduction procedures have also been developed for instruments such as RoboPol \citep{King2014}. These developments demonstrate the growing emphasis on automated, reproducible, and instrument-specific polarimetric data processing frameworks.

Motivated by these advancements, the present work focuses on developing an automated data reduction pipeline for AIMPOL, named DHARA (Data Handling and Automated Reduction pipeline for AIMPOL). 
The pipeline performs end-to-end processing of raw images, including image calibration, source extraction, photometry, computation of Stokes parameters, correction for instrumental polarization, and zero-point offset in angle, and evaluation of final polarization parameters. It is designed to produce science-ready results suitable for further analysis. The structure of the paper is as follows: Sect.~2 provides a brief overview of the telescope, AIMPOL instrument, and data acquisition. Sect.~3 describes the architecture and algorithms of the pipeline; Sect.~4 discusses the results of validation tests; and Sect.~5 summarizes our conclusions and outlines future improvements.

\section{Telescope and instrument}\label{sec:2}
\subsection{Telescope}\label{sec:2.1}
The 104-cm Sampurnanand Telescope \citep{Sampurnanand} is located at Manora Peak, Nainital (latitude \(29^{\circ }22^{\prime }\) N and longitude \(79^{\circ }27^{\prime }\) E) at an elevation of 1951~m. It is a Ritchey–Chrétien reflector operating at the Cassegrain focus and was installed more than 50 years ago by Carl Zeiss, Germany (1972). The telescope is mounted on an equatorial two-pier English mount and has an effective focal ratio of f/13, providing a field of view of approximately 45 arcmin with a Cassegrain-end corrector. The tracking accuracy is about 7 arcsec/hr (0.1 arcsec/min). 
Over the years, the telescope has hosted several imaging detectors, including a Tektronix 1k $\times$ 1k charge couple device (CCD) \citep[first used by][]{Mohan1996}, a 1340 $\times$ 1300 PyLoN CCD \citep{Pandey2023}, and a 4k $\times$ 4k liquid-nitrogen (LN) cooled CCD \citep{4kby4k}. At present, two instruments are in regular operation, the 4k $\times$ 4k LN-cooled CCD imager and AIMPOL, with a significant fraction of observing time dedicated to AIMPOL.
Control of the telescope and its subsystems, such as the secondary mirror focus control, the dome control, and the position control, is semi-automatic using control panels. 

\subsection{Instrument and data acquisition}\label{sec:2.2}
\begin{figure}[!thb]
\includegraphics[width=0.95\columnwidth]{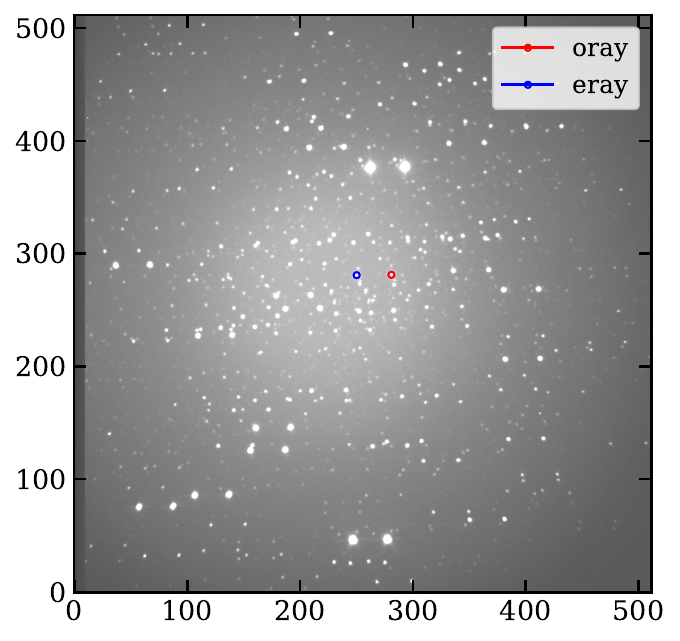}
\caption{Example image obtained with AIMPOL showing the field of the open cluster Koposov~43. Each star appears as a pair of images corresponding to the o-ray and e-ray beams produced by the instrument, and one close to the center is marked by red and blue points, respectively.}\label{Fig:fig1}
\end{figure}

AIMPOL is a back-end instrument at the Cassegrain focus of the 104~cm Sampurnanand telescope. It has been operational since 2004 and is designed to measure the linear polarization of various astronomical objects using the standard Bessel $U$, $B$, $V$, $R$, and $I$ broadband filters. It is a dual-channel optical linear imaging polarimeter, simultaneously measuring two orthogonal polarization states of light coming from a source using a Wollaston prism. The Wollaston splits the incoming light into ordinary (o-ray, $I_o$) and extraordinary rays (e-ray, $I_e$), which are focused on the same detector plane, but with a separation of $\sim 30$ pixels (in the current configuration).  Figure~\ref{Fig:fig1} shows a typical image obtained from AIMPOL, showing the star doublets produced by the instrument, with one example pair highlighted using red and blue markers. 

As described in \citet{Ram1998}, a single measurement in such a system provides one normalized Stokes parameter (either $q$ or $u$). The complementary parameter is obtained by passing the incoming light through an achromatic half-wave plate (HWP), placed before a Wollaston prism on a rotatable mount. In this configuration, a rotation of the HWP by an angle $\alpha$ rotates the plane of polarization by $2\alpha$. Consequently, rotating the HWP by $22.5^\circ$ produces a $45^\circ$ rotation of the polarization plane, allowing the remaining linear Stokes parameter to be determined. Although measurements at HWP positions of $0^\circ$ and $22.5^\circ$ are, in principle, adequate for deriving the linear polarization, observations are generally obtained at four HWP orientations, $\alpha = 0^\circ$, $22.5^\circ$, $45^\circ$, and $67.5^\circ$, to minimize instrumental and CCD-related effects \citep{Ram1998}. The detailed mathematical formalism of the instrument is presented in \citet{Ram1998} and is summarized in \ref{Ap1} for completeness.

The optics of the instrument is currently aligned in such a way that the nominal zeroth position of the HWP corresponds to $67.5^\circ$ and subsequent rotations place the HWP at $45^\circ$, $22.5^\circ$, and $0^\circ$ respectively. Thus, the observations are recorded in the sequence $67.5^\circ, 45^\circ, 22.5^\circ, 0^\circ$, rather than the conventional $0^\circ$, $22.5^\circ$, $45^\circ$, and $67.5^\circ$ order. This configuration has been verified for all our observing epochs from 2017 to 2025.

Earlier, a liquid nitrogen-cooled pylon CCD, with dimensions of $1300 \times 1340$ and a pixel size of $20~\mu m \times 20~\mu m$, was used to detect the light passing through the instrument, giving effective field of view of $\sim 8^{\prime}$, with e-ray and o-ray separation of $\sim 35$ pixels.. Different readout speeds (50, 100, 200, and 500 kHz, and 1, 2, and 4 MHz) and gain (low, medium, and high) options are available in the CCD. The CCD has now been replaced with a $512\times512$ detector having $24~\mu m $ $\times$ $24~\mu m$ pixel size and the same FOV. In the new configuration, the separation between the e-ray and o-ray is approximately $30$ pixels. The corresponding plate scale of the instrument is $1.43~\mathrm{arcsec/pixel}$, and a representative radial profile of a stellar image obtained with the current configuration is shown in \ref{Ap:A}. Further details of AIMPOL are presented in \citet{Rautela2004}, with an upgraded version described in \citet{Pandey2023}. The instrument has been extensively used to study ISM polarization, comets, Blazars, AGNs, supernovae, and GRBs \cite[e.g.,][]{Soam2017, Choudhury2022,  Uppal2024, Lee2018, Dastidar2019, Pandey2021}.                    

\section{Description of pipeline}\label{sec:3}
\textsc{DHARA} pipeline is specifically designed to reduce and analyze data from the AIMPOL instrument. The entire pipeline is written in Python~3 programming language using the easily available Python packages like \textsc{scipy}, \textsc{astropy}, \textsc{photutils}, and \textsc{astroquery}. It is an interactive-automatic pipeline that takes the path/ file name of the data to process raw images saved in a directory, calibrate them, and provide science-ready results, including sky position (right ascension, RA and declination, DEC), Stokes parameters, degree of polarization, and polarization angle, along with their $1\sigma$ uncertainties. The pipeline operates in a largely automated manner, and processes raw frames stored in FITS format. The AIMPOL FITS headers currently include basic observational metadata such as exposure time, detector configuration, pixel array information, and acquisition date. However, key parameters for polarimetric data reduction, such as the filter used, the HWP orientation, and the sky coordinates, are not recorded in the header. Consequently, for instruments such as AIMPOL, the pipeline requires a small set of user-defined configuration parameters specifying the filter and the mapping between exposure identifiers and corresponding HWP angles. To ensure reliable association of each frame with its instrumental configuration throughout the reduction process, it is recommended that these missing metadata be explicitly encoded within the file naming convention (e.g., we use the name convention as \texttt{object\_filter\_exptime\_HWP\_set.fits} during our observations). 

\begin{figure}[!thb]
\includegraphics[width=0.95\columnwidth]{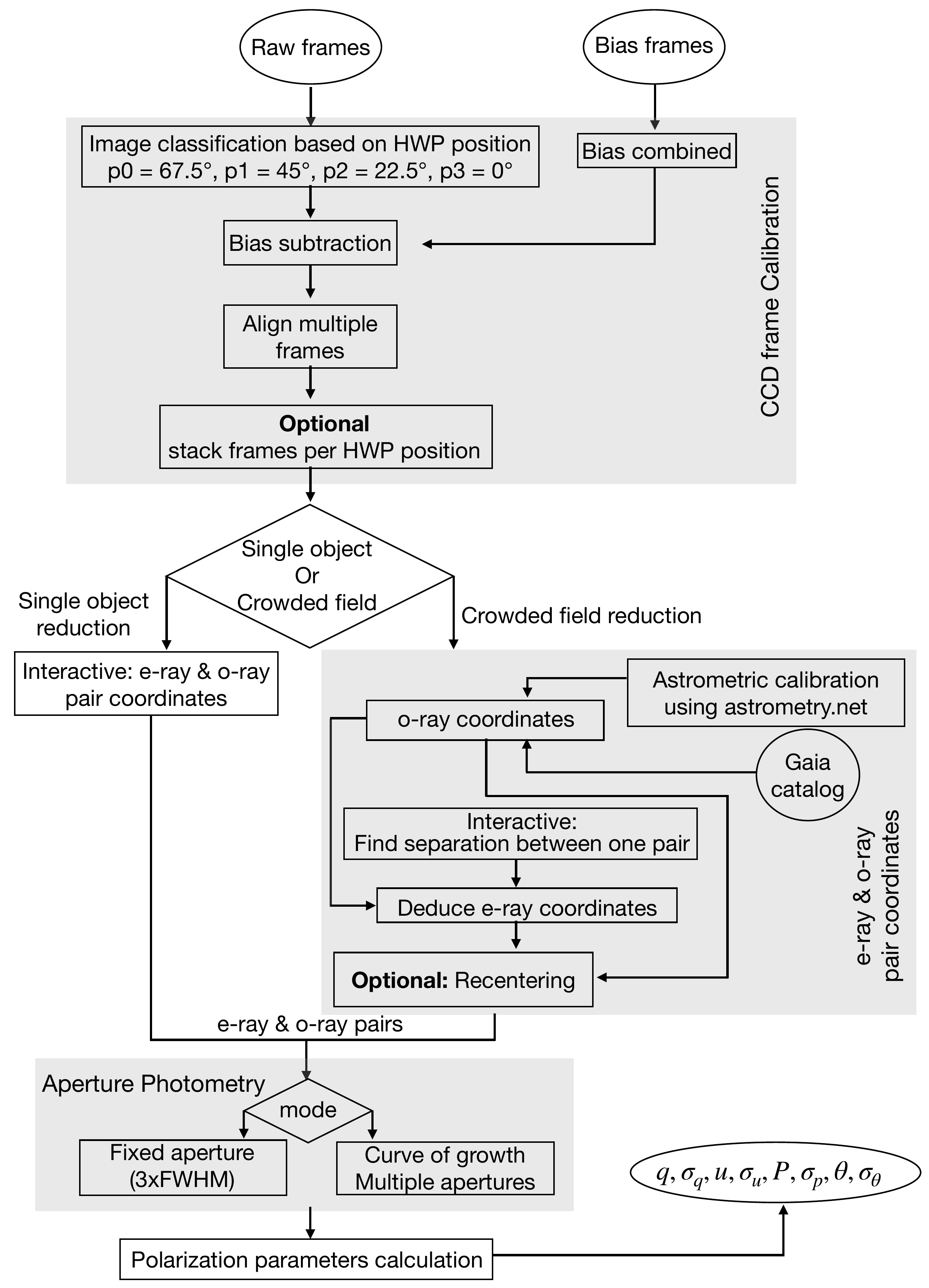}
\caption{Schematic recipe for reduction procedure implemented in \textsc{DHARA} pipeline.}\label{Fig:chart}
\end{figure}

The initial tasks involved in the reduction are bias subtraction, shifting, and stacking of multiple sets, and source extraction. We utilize the web-based \textsc{astrometry.net} \citep{Lang2010} service to obtain the sky coordinates of the extracted stars in crowded fields. Separating the e-ray and o-ray images of each star is a crucial step to perform relative aperture photometry, which is used to compute the Stokes parameters and, hence, the degree of linear polarization and polarization angle. A schematic flowchart of the pipeline is shown in Fig.~\ref{Fig:chart}. The detailed descriptions of the different tasks are provided in the following subsections.
\subsection{CCD frame calibration}\label{sec:3.1}
The basic CCD calibration procedure begins with the construction of a master bias frame. All bias (zero-exposure) images were median-combined to generate the master bias, which is then subtracted from each science frame. Flat-field correction is not required for AIMPOL observations, as the pixel-to-pixel sensitivity variations are accounted by the factors $F_e$ and $F_o$ defined in Eqs.~\ref{eq4}, \ref{eq5}, and \ref{eq6}.

For each science target, the raw images are first classified according to the HWP position using the identifier \textit{`HWP', `HWPANG', `HWP\_POS' in the FITS header or} encoded in the filename. In current implementation, the HWP orientations are denoted by p0, p1, p2, and p3 in the filename and correspond to $67.5^\circ$, $45^\circ$, $22.5^\circ$, and $0^\circ$, respectively. This classification ensures that frames corresponding to the same polarization state are grouped together prior to further processing.

In practice, multiple exposures with identical integration times\footnote{As linear polarization is estimated from relative intensity measurements at different HWP orientations, exposures at all HWP positions are required to have identical integration times to maintain consistent flux scaling across frames.} are typically acquired at every HWP position. These individual exposures are generally distinguished through filename identifiers such as s1, s2, s3, or alternatively by numerical suffixes (e.g., 1, 2, 3). The exposures corresponding to a given HWP position can subsequently be combined to improve the signal-to-noise ratio (SNR). However, small frame-to-frame shifts introduced by imperfect telescope tracking may be present. While these shifts are typically negligible in short exposures, they can become significant in longer integrations. Therefore, prior to stacking, it is crucial to align them to a common reference frame by ensuring consistent centering. 

Image alignment for each set is performed iteratively using \texttt{find\_transform} and \texttt{apply\_transform} routines of \textsc{astroalign} Python library. This procedure ensures that all frames are registered to a common reference frame through small geometric transformations. However, \textsc{astroalign} may fail in cases involving large relative shifts between frames. In such situations, the alignment is instead carried out by estimating the relative shifts between frames using the centroid of a bright, isolated reference star. The centroid in the successive images are determined with the  \texttt{centroid\_quadratic} algorithm in \textsc{photutils}. The images are then shifted to a common reference position using the \texttt{shift} routine from the \texttt{ndimage} module of the \textsc{scipy} library. In this process, the pixel values in the non-overlapping area are set to be zero to avoid any confusion. The aligned frames corresponding to each HWP position can be median or sum combined  to produce stacked images with improved signal quality as an optional step.

\subsection{Astrometric calibration}\label{sec:3.2}
Scientific analysis of the CCD images requires accurate sky coordinates for the detected sources, particularly in crowded fields. However, the raw CCD images do not contain any information on sky coordinates. We can only obtain the pixel coordinates of the stars in the image, which also can be a tedious task if done manually, particularly when dealing with crowded fields. To obtain an astrometric solution, we used the web-based \textsc{astrometry.net} software \citep{Lang2010}, which provides a transformation between image coordinates and sky coordinates.

As described in the Sect.~\ref{sec:2.2}, the o-ray and e-ray images of each star are recorded on the same CCD with a separation of $\sim 30$ pixels; the astrometry service provides a solution for only one component of each pair (either the o-ray or the e-ray set). 
 However, for performing our next step, which is the relative photometry on e-ray and o-ray pair, we require the image coordinates of all pairs in the field. For a particular star in a field or isolated sources, such as standard stars, transients, or fields where an astrometric solution cannot be reliably obtained through Astrometry.net, the e-ray and o-ray coordinates are determined interactively using `single star' option. However, for a crowded field, we first determined the image coordinates for one set of stars for which astrometric solution is available. We select the RA, DEC coordinates of sources in the field from the Gaia DR3 catalog down to a user-defined limiting magnitude and convert their sky positions to image coordinates based on astrometric solution using the \textsc{astropy} \texttt{WCS} module. Since the separation between the o-ray and e-ray sources is fixed (within $\pm1$ pixel, see \ref{Ap:C} for details) and can be determined by identifying a single o-ray and e-ray pair in the field, we can easily obtain the coordinates of the other set of stars (o-ray or e-ray) by applying the fixed separation. Thus, the image coordinates of all the e-ray and o-ray source pairs are determined along with their sky coordinates.

We provide an optional recentering step to account for residual astrometric uncertainties and potential errors in the determination of the separation between o-ray and e-ray images, which can cause small shifts in the detected pixel positions of each pair. For this purpose, the centroids of the o-ray and e-ray images are recomputed using their initially derived pixel coordinates as starting points, employing the \texttt{centroid\_quadratic} and \texttt{centroid\_2dg} routines from the \texttt{centroid} package of \textit{photutils}. 
\subsection{Photometry}\label{sec:3.3}
The computation of polarization measurement requires relative intensities of e-ray to o-ray at each HWP position. Once the accurate pixel coordinates of all the stars in the field are obtained, aperture photometry is performed utilizing the \textsc{photutils} module of Python. The local background is estimated from the median signal in an annular aperture surrounding the star and subtracted from the source counts.
A constant optimal aperture corresponding to three times the full width at half maxima ($3 \times$ FWHM) is used to carry out the photometry of both the o-ray and e-ray images of a star. 

However, in the crowded field, there is a finite probability for the aperture to overlap the nearby stars. The crowding problem in AIMPOL images is multiplied because e-ray and o-ray images are present together, hence the e-ray of one star may overlap the aperture of the o-ray of the other, and vice-versa, adding to the complexity of the photometric analysis \citep[see Fig.~4 of][]{uppal2023optical}.  To minimize this effect, we have an option of using multiple apertures, defined between 1 $\times$ FWHM and 3 $\times$ FWHM, and the photometric counts are computed in each aperture for each HWP position. An annulus with fixed inner and outer radii is used to calculate the median background and subtract the background from the corresponding aperture. The aperture size that gives the lowest error in the polarimetry, or the aperture at which the polarization curve of growth starts saturating, is selected to compute the final polarization parameters of a star (described in the Sec .~\ref {sec:3.4}).

\subsection{Polarimetery}\label{sec:3.4}
\begin{figure}[!thb]
\includegraphics[width=0.95\columnwidth]{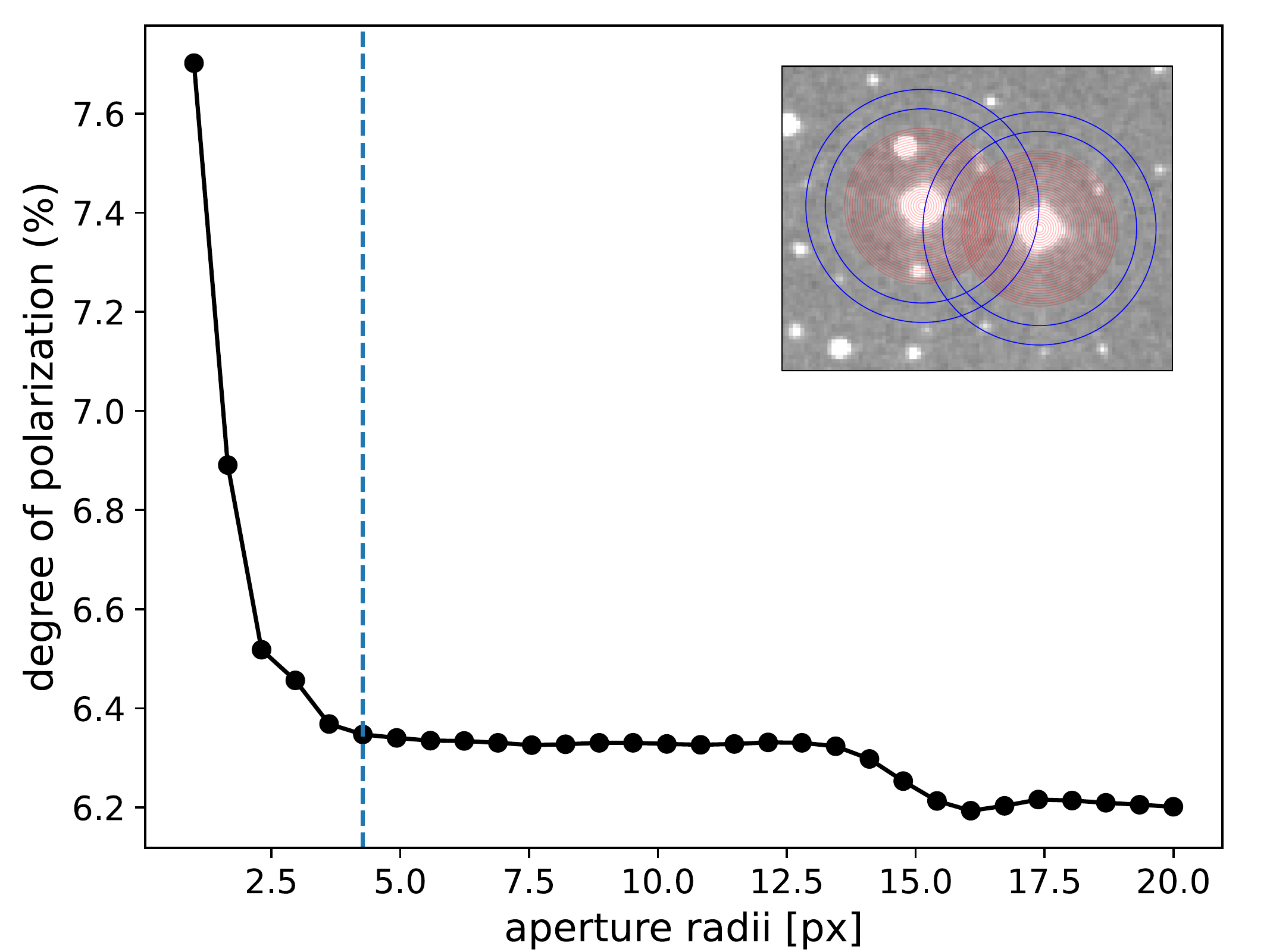}
\caption{Polarization curve of growth for HD 344776, showing the degree of polarization as a function of increasing aperture size. The optimal aperture is marked by a blue dashed line. The inset shows a cut-out image containing the e-ray and o-ray components of HD 344776. Multiple circular apertures used for photometry are shown in red, while the background annulus is indicated by blue circles.}\label{Fig:aper}
\end{figure}

As discussed in \ref{Ap1}, the polarization calculation depends on the ratio of measured o-ray ($I_o^\prime$) and e-ray ($I_e^\prime$) counts at different HWP positions. The modulation factor from eq.~\ref{eq3} and eq.~\ref{eq4}, \ref{eq5}, \ref{eq6} is modified as 
\begin{equation}\label{eq7}
     R(\alpha) = \frac{\frac{I_e^\prime}{I_o^\prime} \times \frac{F_o}{F_e}  -1}{\frac{I_e^\prime}{I_o^\prime} \times \frac{F_o}{F_e} +1} 
\end{equation}
and the uncertainty in the measurement of R$(\alpha)$ in low polarization regime \citep{Ram1998} is approximated to
\begin{equation}\label{eq8}
     \sigma_{R(\alpha)} = \frac{\sqrt{I_e + I_o + 2I_b}}{I_e + I_o},
\end{equation}
where $I_b$ is the average background counts around the e-ray and o-ray images.
Using $q = p\cos{2\theta}$ and $u = p\sin{2\theta}$, the right side of the eq.~\ref{eq3} can be re-written as 
\begin{equation}\label{eq9}
    R(\alpha) = q_{obs}\cos{4\alpha} + u_{obs}\sin{4\alpha}. 
\end{equation}
This modified $R(\alpha)$ (eq.~\ref{eq7}), for all the apertures, is fitted with the eq.~\ref{eq9}. The aperture radius at which the polarization curve of growth begins to saturate is then adopted to derive the final polarization parameters of the source, namely the polarization fraction and polarization angle. Deviations from this stable polarization value at larger apertures indicate the inclusion of additional flux components, i.e., the presence of a nearby star. The curve-of-growth method, therefore, provides a means to select an optimal aperture that minimizes contamination from nearby sources. An example of this procedure is shown for the star HD~322776 in Fig.~\ref{Fig:aper}. The figure presents the polarization curve of growth, while the inset shows the o-ray and e-ray images of the star with the apertures marked by red circles and the background annulus in blue. The vertical dashed line marks the optimal aperture, corresponding to the onset of saturation in the polarization curve. The subsequent variation in polarization at larger aperture radii highlights the effect of flux contamination from nearby sources.

Sources whose apertures overlap with one another are further examined. For such cases, polarization measurements are reported using the smallest aperture corresponding to $1\times$ FWHM, provided that the sources do not overlap at this aperture. If they remain overlapping even at $1\times$ FWHM the sources are flagged and excluded from subsequent analyses. 

The obtained Stokes parameters of all the observed stars ($q_{obs}$ and $u_{obs}$) and their uncertainties ($\sigma_{q_{obs}}, \sigma_{u_{obs}}$) are corrected for instrumental polarization, in the $qu-$space following eqs.~\ref{eq10} and \ref{eq11}.
\begin{equation}\label{eq10}
    q = q_{obs} - q_{inst} \;, \\
    u = u_{obs} - u_{inst} \; , 
\end{equation}

\begin{equation}\label{eq11}
    \sigma_q = \sqrt{\sigma_{q_{obs}}^2 + \sigma_{q_{inst}}^2} \;,\\
     \sigma_u = \sqrt{\sigma_{u_{obs}}^2 + \sigma_{u_{inst}}^2} \; ,
\end{equation}
Where, ($q_{inst}, u_{inst}$, $\sigma_{q_{inst}}$, and  $\sigma_{u_{inst}}$ ) are the Stokes parameters and their uncertainties corresponding to instrumental polarization. 

In addition, any zero-offset in the polarization angle ($\psi_{inst}$), determined by comparing measurements of polarized standard stars with their literature values, is corrected so that all angles are reported in a common reference frame, i.e., measured from celestial north and increasing toward the direction of right ascension according to IAU conventions. For both $q$ and $u$, the correction can be applied in a single step using the following equation to find the final Stokes parameters ($q, u$)
\begin{equation}\label{eq12}
    \begin{bmatrix} q_{obs} \\ u_{obs} \end{bmatrix} = \begin{bmatrix} q_{inst} \\ u_{inst} \end{bmatrix} + \begin{bmatrix} \cos{(2\psi_{inst})} & \sin{(2\psi_{inst})} \\ -\sin{(2\psi_{inst})} & \cos{(2\psi_{inst})} \end{bmatrix} \begin{bmatrix} q \\ u \end{bmatrix} 
\end{equation}

The instrumental parameters, $q_{inst}$, $q_{inst}$, and $\psi_{inst}$ are used as preset values in the pipeline. They are not fixed a priori in the pipeline but need to be determined for each observing run using unpolarized and polarized standard stars.  The unpolarized stars are first processed with these terms initially set to zero to estimate $q_{inst}$, $u_{inst}$ for that epoch. These values are then input into the reduction of polarized standards and the resulting polarization angle are compared with the literature values to determine the average zero-angle offset, $\psi_{inst}$. 

Once determined, $q_{{inst}}$, $u_{{inst}}$, $\psi_{{inst}}$, along with their associated uncertainties, are supplied as preset parameters to the pipeline for the final reduction of all science targets. The pipeline then applies Eq.~\ref{eq12} to obtain the Stokes parameters of a star, from which the degree of polarization and polarization angle are derived as follows.
\begin{equation}\label{eq13}
    p_{obs} =  \sqrt{q^2 + u^2} \; ,
\end{equation}
and the polarization angle is determined as: 
\begin{equation}\label{eq14}
  \theta = \frac{1}{2}\; {\rm{arctan2}}(u,\,q) \;.
\end{equation}
Here, we use the two-argument arctangent function to define polarization angle in order to account for the inherent $180^\circ$ ambiguity. 

The propagated uncertainties in the derived degree of polarization (Eq.~\ref{eq13}) and polarization angle (Eq.~\ref{eq14}) are estimated using 
using:
\begin{equation}\label{eq15}
 \sigma_{p} =  \sqrt{\frac{q^2\sigma_{q}^2 + u^2\sigma_{u}^2}{q^2+u^2}} \;,
\end{equation}
\begin{equation}\label{eq16}
 \sigma_{\theta} = \frac{1}{2(q^2+u^2)} \sqrt{q^2\sigma_{u}^2 + u^2\sigma_{q}^2} \; .
\end{equation}
The degree of polarization derived from Stokes parameters (Eq.~\ref{eq13}) follows a Ricean distribution.
The uncertainties in $q$ and $u$, arising from multiple instrumental and observational sources, propagate in a positive-definite manner and therefore tend to systematically increase the estimated degree of polarization \citep{Patat2006, Sohn2011}. This effect introduces a bias in the measured polarization, particularly when the signal-to-noise ratio is low, and the uncertainty becomes comparable to the polarization signal itself. To mitigate this Ricean bias in low-SNR regimes, we apply the asymptotic estimator \citep{wardle1974}, expressed as:
\begin{equation}\label{eq17}
    P = \sqrt{p_{obs} ^2 - \sigma_{p} ^2}\;.
\end{equation}
In contrast, the polarization position angle follows a different statistical distribution compared to the degree of polarization \citep{Naghizadeh1993}. Since $q$ and $u$ share similar noise characteristics, their ratio, which determines the position angle (see Eq.~\ref{eq14}), is largely insensitive to this bias. As a result, the position angle is expected to be significantly less affected by statistical bias than the degree of polarization. Therefore, no debiasing correction is applied to the polarization angle.
However, the uncertainty in the angle is biased and it is estimated following the formalism of \cite{Naghizadeh1993}, which accounts for the non-Gaussian nature of polarization angle at low signal-to-noise ratio as described in \cite{Blinov2023}. Hence, to evaluate the corrected uncertainties in the polarization angle, we followed the same procedure outlined in \cite{Blinov2023}. 

\section{Results and validation}\label{sec:4}
To assess the reliability of the results produced by the \textsc{DHARA} pipeline, we compared our measurements with published literature values for both isolated sources and crowded fields. For the single-source validation, we observe $26$ polarized standard stars from AIMPOL across multiple epochs. To evaluate the performance of pipeline in crowded environments, we applied the our pipeline to observations of a Galactic open cluster field. 

\subsection{Standard stars}\label{sec:4.1}
The analysis of polarized standard stars is essential for validating the accuracy of our measurements. We observed several polarized standard stars spanning a wide range of degrees of polarization from as low as $0.2\%$ to as high as $6.5\%$, across multiple observing epochs between 2017 and 2025 in the R-band. 
For the 26 standard stars listed in Table~\ref{tab:pol_appendix} of \ref{Ap:D}, multiple sets of polarimetric observations are obtained over different epochs and processed using our data reduction pipeline. Instrumental polarization is corrected using observations of zero-polarization standard stars\footnote{The unpolarized standards used in different epochs include G191B2B, HD~14069, BD+28$^\circ$4211, HD~212311, BD+29$^\circ$2198, BD+35$^\circ$2256, and GD~319, selected from the compilations of \citet{Schmidt1992} and references therein.} observed each night in $qu-$space. As all targets are relatively bright (Gaia Gmag ranging from 4.74 to 13.50~mag), each observational set is reduced independently. We determine the weighted mean Stokes parameters $q$ and $u$ of independent sets, from which the degree of polarization and polarization angle were derived. 
The associated uncertainties are propagated from the errors on the weighted means of $q$ and $u$ using Eqs.~\ref{eq15} and \ref{eq16}. The resulting polarization measurements are then compared with corresponding values reported in the literature.

Table~\ref{tab:pol_appendix} lists the polarized standard stars (column~2), their observation dates (column~1), the computed instrumental corrected degree of polarization ($P_{\rm{pipe}}$) and polarization angle ($\theta_{\rm{pipe}}$) along with their uncertainties ($\sigma_{p_{\rm{pipe}}}$, $\sigma_{\theta_{\rm{pipe}}}$) (in columns~3 and 4), and the corresponding literature values ($P_{\rm{lit}}$, $\sigma_{p_{\rm{lit}}}$, $\theta_{\rm{lit}}$, and $\sigma_{\theta_{\rm{lit}}}$ in columns~5 and 6). Column~7 represents the zero-angle offset between the observed and literature value, expressed in range $0^\circ–180^\circ$. The reference of the literature values are given in column~8. 
Although standard stars are used to determine the zero-angle calibration, the telescope orientation relative to the sky and the waveplate positions within the polarimeter vary between observing campaigns. Consequently, the offset angle ($\psi_{inst}$, column~7) differs between epochs and cannot be meaningfully compared among stars observed at different times. Nevertheless, for observations obtained within the same observing epoch, the measured offset angles remain consistent within the uncertainties for all stars except L\_98\_685. In our observations from 07 March 2018 and 08 March 2018, the measured polarization angles for L\_98\_685 deviate from the literature values as the derived $\psi_{inst}$ is significantly different from that of the other standard stars observed during the same epoch. Since only a single epoch of observations for this star is reported in \citet{Blinov2023}, its long-term polarization behavior remains uncertain. The observed discrepancy may therefore indicate possible intrinsic variability, which would require further multi-epoch observations to confirm or low polarization SNR effect.

\begin{figure}[!thb]
\includegraphics[width=0.95\columnwidth]{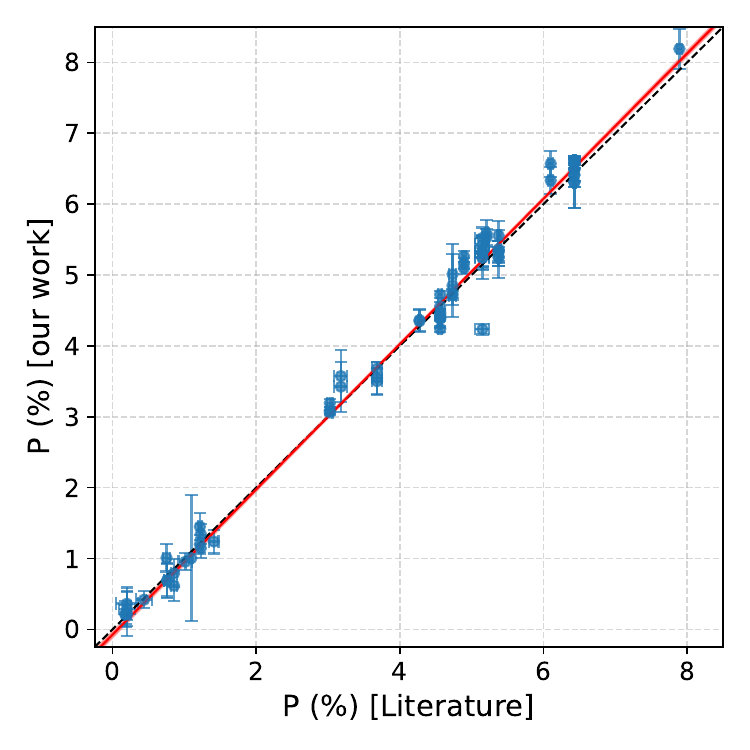}
\caption{Comparison of the degree of polarization of polarized standard stars data reduced from our pipeline with the literature values.  The black dashed line indicates the one-to-one correspondence (x = y). The red line represents the ordinary least square regression fit to the data.} \label{Fig1}
\end{figure}

The measured degree of polarization of all the observed stars shows good agreement with the literature value within $2\sigma$ with the exception of HD~251204.
The literature value reported for HD~251204 in Table~\ref{tab:pol_appendix} corresponds to the V band, whereas our observations were obtained in the R band. Figure~\ref{Fig1} shows a comparison between the degrees of polarization derived from our pipeline and the corresponding literature values for the standard stars listed in Table~\ref{tab:pol_appendix}. Star HD~251204 is excluded from this plot because the observed values and the literature values correspond to different filters.   Additionally, this source may exhibit intrinsic polarization variability \citep{Amirkhanyan2006}, which could contribute to the observed discrepancy. We also note that the polarization uncertainties estimated by the our pipeline are larger than those reported in the literature, likely due to the reduction of statistical noise in previous studies through extensive repeated observations.
Overall, the observed degrees of polarization for the standard stars show good agreement with the literature values, as indicated by the $x = y$ dashed line in the figure. We performed a linear fit of the form $p[\rm{our work}] = m~p[\rm{literature}] + c$, using orthogonal distance regression. The best-fit parameters for slope and intercept are $m = 1.02 \pm 0.01$ and $c = -0.08 \pm 0.04$, respectively, shown by the red line in the plot, indicating consistency between the two datasets within the $2\sigma$ level.

\subsection{Open cluster}\label{sec:4.2}
\begin{figure}[!t]
\includegraphics[width=0.95\columnwidth]{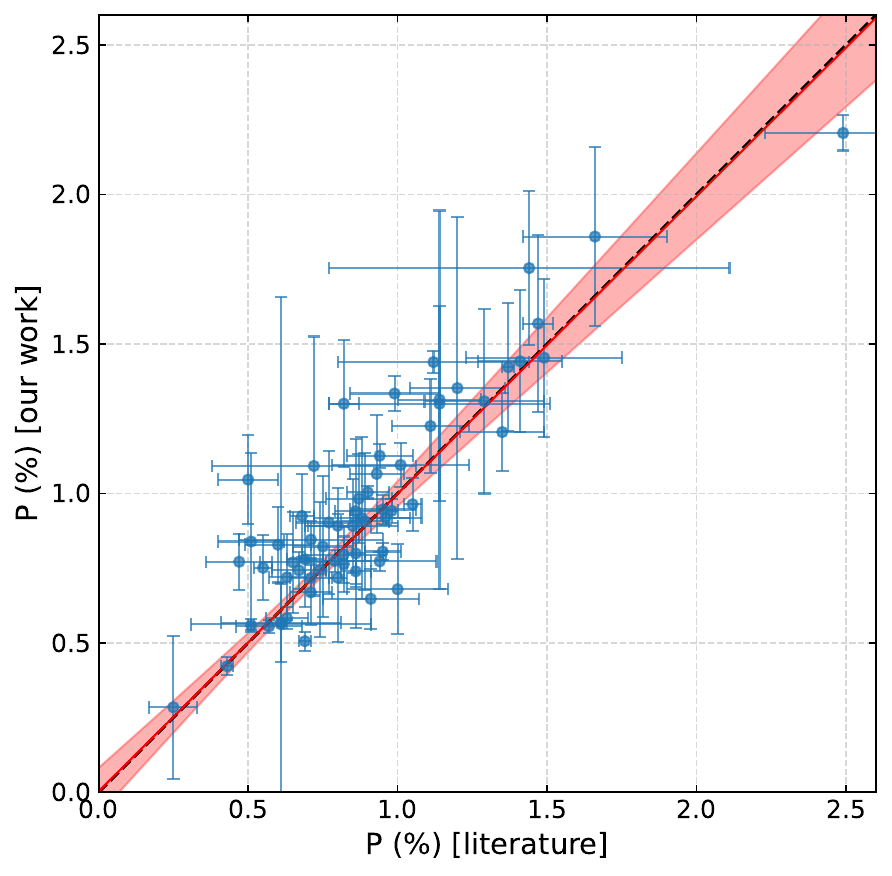}
\caption{Comparison between the degree of polarization of stars toward the Alessi~1 open cluster derived using our pipeline and those reported by \citet{Sadhana2020}. The dashed line indicates the one-to-one correspondence (x = y) and red line and shaded region corresponds to the ordinary least square fit to the data.}\label{Fig:2}
\end{figure}

As discussed in Sect.~\ref{sec:3.3}, photometry and, consequently, polarization estimation become challenging in crowded fields, particularly for a dual-channel polarimeter in which the e-ray and o-ray images of each source are present on the same CCD. To validate the reliability of our pipeline under such conditions, we reanalyzed polarization data of the open cluster Alessi~1, previously studied by \citet{Sadhana2020}. In their work, astrometry and aperture photometry were performed using the standard Image Reduction and Analysis Facility (IRAF) to obtain the fluxes of the e-ray and o-ray images by manually selecting each isolated star. The same dataset is processed with our pipeline to derive the polarization parameters for their set of stars. We find a small systematic zero-point offset of approximately 0.05\% in the degree of polarization relative to the published values. Such a global offset can plausibly arise from differences in the estimation and correction of instrumental polarization between independent analyses. 
To allow a direct comparison on a common zero point, a uniform shift of 0.05\% was applied to all polarization measurements derived in this work and is presented in Fig.~\ref{Fig:2}. An ordinary linear least-squares regression (represented by the red line and spread in the shaded region) yields a slope of $0.99 \pm 0.05$ and an intercept of $0.004 \pm 0.039$, consistent with the one-to-one ($x=y$) relation within $1\sigma$. The close agreement within $1\sigma$ uncertainties demonstrates that our measurements are consistent with the literature and validates the accuracy of our pipeline for polarization measurements in crowded fields.

\section{Summary and conclusion}\label{sec:5}
We have presented \textsc{DHARA}, an automated data-reduction pipeline developed for imaging polarimetry using dual-channel polarimeters such as AIMPOL. The pipeline processes the raw images from the instrument, identifies e-ray and o-ray pairs, performs aperture photometry on them, and computes polarization parameters corrected for instrumental polarization and zero angle offset. It produces a final science-ready catalog containing the sky coordinates of detected sources, Stokes parameters, the corresponding degree of polarization, and polarization angle, along with their associated $1\sigma$ uncertainties as output.

To validate the performance and reliability of the pipeline, we analyzed the polarization observation data for both isolated and crowded fields. This included a sample of 26 polarized standard stars and Alessi~1 open cluster field. In both cases, the polarization measurements derived using our pipeline show good agreement with published literature values, consistent within $2\sigma$ uncertainties. In its current implementation, the pipeline employs aperture photometry; however, we must exclude stars affected by overlapping apertures in crowded regions for further analyses. To address this limitation, PSF photometry will be included in a future version of the pipeline. 

A key advantage of our pipeline is its fully automated workflow, which significantly reduces the time required for data reduction and analysis. While manual selection and processing of sources can be time-consuming, \textsc{DHARA} generates a complete polarization catalog within minutes. The pipeline is broadly applicable to any dual-channel polarimeter in which the e-ray and o-ray images are recorded on the same CCD plane with a small spatial separation.                                       
\section*{Acknowledgements}
We thank the anonymous reviewers for their comments and valuable suggestions, which have improved the clarity and quality of the manuscript. We acknowledge the Telescope Time Allocation Committee of the Sampurnanand Telescope (ARIES) for approving and allocating observing time for our proposals. We thank the local staff,  colleagues at the Sampurnanand Telescope, and the observers for their assistance during the observations. This work was initiated at the Physical Research Laboratory, which is supported by the Department of Space, Government of India. We thank Prof. A. N. Ramaprakash, Inter-University Centre for Astronomy and Astrophysics (IUCAA), for the helpful discussions.
N.U. acknowledges support from an European Research Council (ERC) grant, MW-ATLAS project no. 101166905. N.U. and D.B. acknowledge support from the Horizon ERC Grants 2021 programme under the grant agreement No. 101040021.
N.U. gratefully acknowledges the computational facilities and office space provided by IUCAA, which enabled the completion of this work. 

\noindent
\textbf{Sofware:} Astropy \citep[\href{https://www.astropy.org/}{https://www.astropy.org/})][]{Astropy2013}, Scipy (\href{https://scipy.org/}{https://scipy.org/}), Photutils (\href{https://photutils.readthedocs.io/en/stable/}{https://photutils.readthedocs.io/en/stable/}) \citep{Bradley2016}, Astrometry.net \citep[\href{https://astrometry.net/}{https://astrometry.net/})][]{Lang2010}
\vspace{-1em}

\bibliographystyle{jaa}
\bibliography{ref} 


\appendix
\section{Mathematical formalism}\label{Ap1}
Following the mathematical formalism for the instrument outlined in \citep{Ram1998}, the intensities recorded in the extraordinary and ordinary beams ($I_e$ and $I_o$) can be expressed as
\begin{equation}\label{eq1}
    I_e (\alpha) = \frac{I_{up}}{2} +  I_p \cos^2{(\theta - 2\alpha)}
\end{equation}
\begin{equation}\label{eq2}
    I_o (\alpha) = \frac{I_{up}}{2} +  I_p \sin^2{(\theta - 2\alpha)}
\end{equation}
where $I_{up}$, and $I_p$ are the unpolarized and polarized intensities, respectively. The angles $\theta$ and $\alpha$ are the position angles of the polarization vector and the HWP fast-axis, respectively, both defined with respect to the reference axis of the Wollaston prism. 

A modulation factor $R(\alpha)$ is then defined as
\begin{equation}\label{eq3}
    R(\alpha) = \frac{I_e/I_o -1}{I_e/I_0 +1} = p \cos(2\theta - 4\alpha)
\end{equation}
where, $p = \frac{I_p}{I_p + I_{up}}$ denotes the degree of linearly polarization expressed in fraction. For HWP orientation, $\alpha = 0^\circ$ and $22.5^\circ$ this expression reduces to the normalized Stokes parameters $Q/I$ and $U/I$, respectively.

In practice, the HWP is rotated at four position angles, $\alpha = 0^\circ$, $22.5^\circ$, $45^\circ$, and $67.5^\circ$ \citep{Ram1998}.This approach is adopted to mitigate systematic effects arising from possible differences in the system response to e-ray and o-ray polarization states, and spatial variations in CCD sensitivity across the detector field. 
As a result, the measured intensities in the two beams, denoted by $I_e^\prime$ and $I_o^\prime$, are modified by instrumental response factors $F_e$ and $F_o$, such that
\begin{equation}\label{eq4}
    I_e^\prime = I_e \times F_e
\end{equation}
and
\begin{equation}\label{eq5}
    I_o^\prime = I_o \times F_o.
\end{equation}
The ratio of these response factors can be calculated as 
\begin{equation}\label{eq6}
    \frac{F_o}{F_e} = \left[\frac{I_o^\prime (0^\circ)}{I_e^\prime(45^\circ)} \times \frac{I_o^\prime (45^\circ)}{I_e^\prime(0^\circ)} \times \frac{I_o^\prime (22.5^\circ)}{I_e^\prime(67.5^\circ)} \times \frac{I_o^\prime (67.5^\circ)}{I_e^\prime(22.5^\circ)}\right]^{1/4}
\end{equation}
making use of the fact that a rotation of the half-wave plate by $45^\circ$ effectively interchanges the ordinary and extraordinary beams. Consequently, to compensate for CCD effects, measurements are obtained by rotating the HWP by $45^\circ$ and $67.5^\circ$ in addition to $0^\circ$ and $22.5^\circ$. 

\section{Radial profile of star in AIMPOL}\label{Ap:A}
\begin{figure}[!h]
\includegraphics[width=0.94\columnwidth]{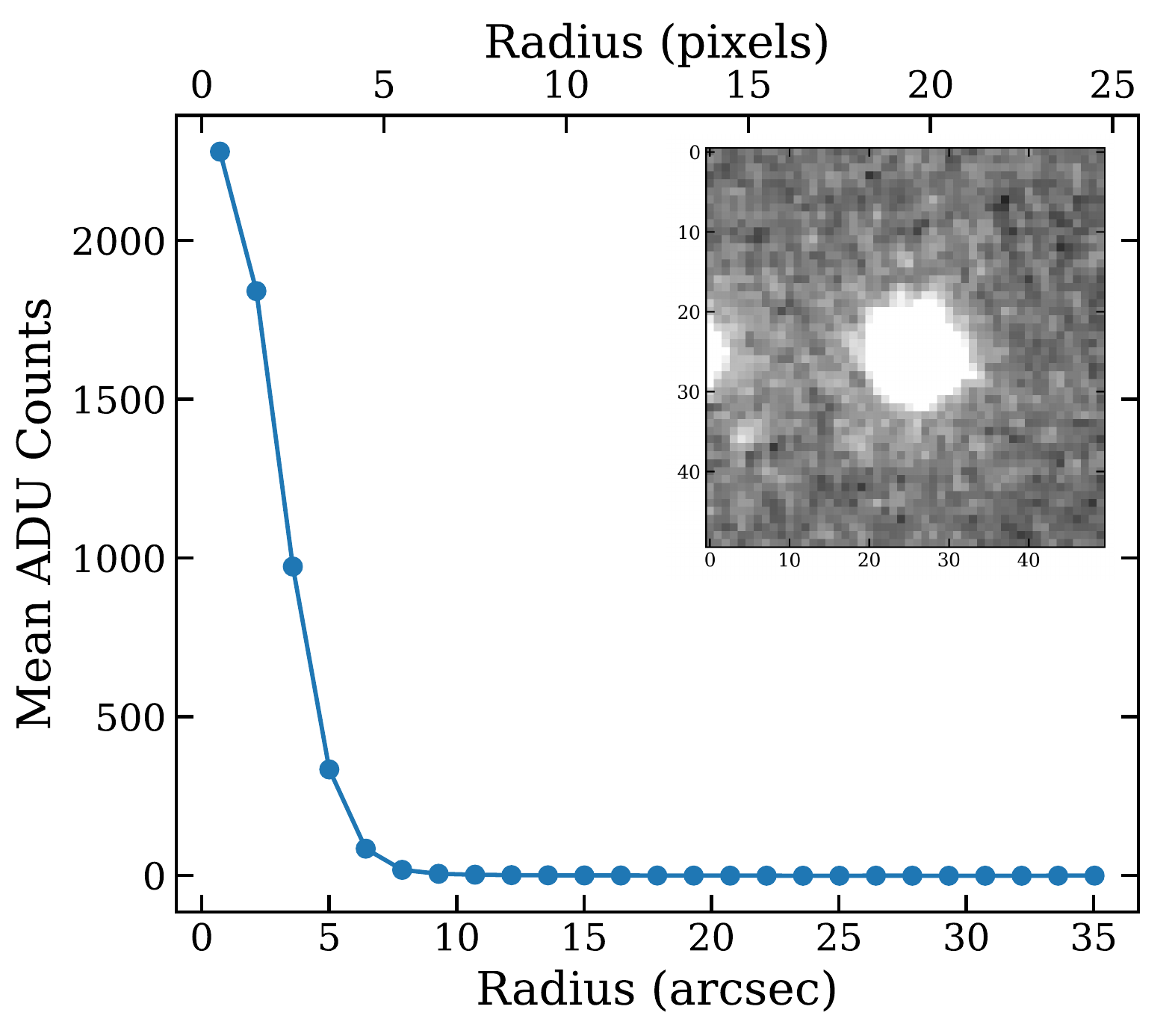}
\caption{Radial profile and a cut-out (as an inset on the top-right corner) of a o-ray image of a star, BD+59~389 in AIMPOL field.}\label{Fig:C1}
\end{figure}

As discussed in Section~\ref{sec:2.1} AIMPOL instrument with its current $512\times512$ CCD configuration has a plate scale of $1.43~\mathrm{arcsec/pixel}$. Since the detector sampling and image quality influence the reliability of the photometric measurements, we analyzed the radial profile of a star image recorded on CCD. Fig~\ref{Fig:C1} shows the radial profile of o-ray image of the star BD+59~389, with the corresponding image cutout displayed as an inset of the figure, located at the top-right corner. The profile yields an FWHM of $\sim 3.55$ pixels, indicating that the PSF is well-sampled by the detector based on the Nyquist criterion and is suitable for accurate centroid and photometric analysis.   

\section{Quantitative assessment of o- and e-ray separation across CCD}\label{Ap:C}
In the current instrument configuration with $512\times 512$ CCD, the e-ray and o-ray images are separated by $\sim 30$ pixels. To validate the assumption of a constant separation across the field (section~\ref{sec:3.2}), we performed an independent test using an open cluster image, which provides a dense field and spatially samples the entire CCD area.
 
\begin{figure}[!htb]
\includegraphics[width=0.95\columnwidth]{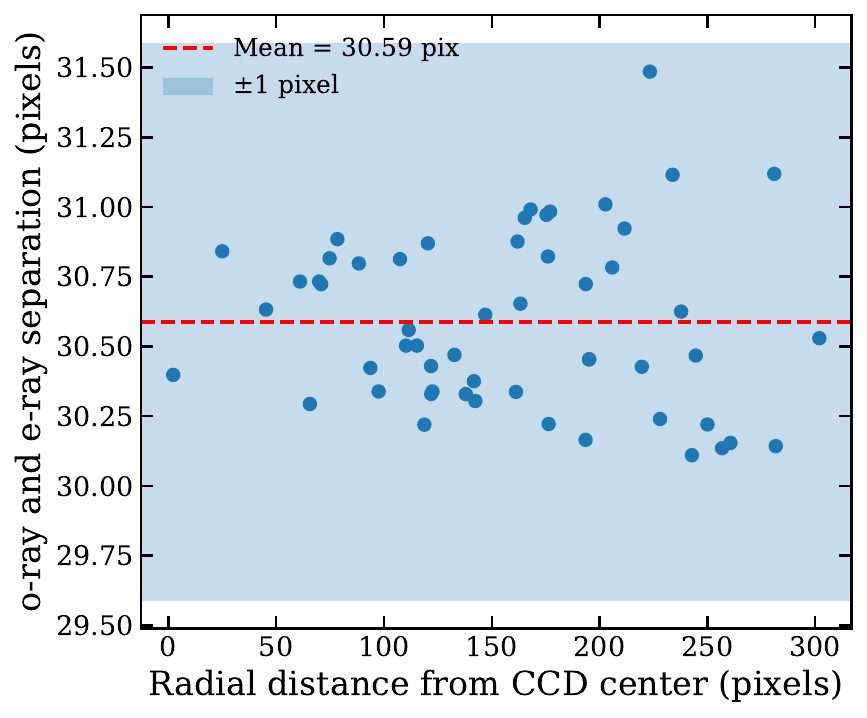}
\caption{Separation between e-ray and o-ray pair as a function of radial distance from CCD center from an AIMPOL field observed towards the NGC2269 cluster.}\label{Fig:B1}
\end{figure}

\begin{figure}[!htb]
\includegraphics[width=0.95\columnwidth]{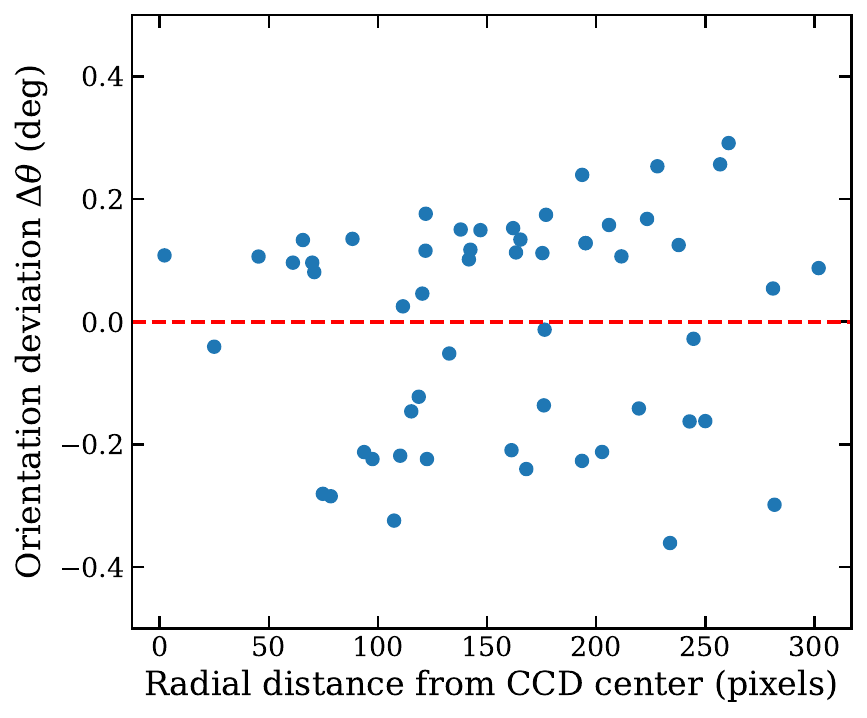}
\caption{Orientation deviation of e-o image pair from their mean value as a function of radial distance from CCD center.}\label{Fig:B2}
\end{figure}

We selected non-overlapping o-ray and e-ray image pairs at different positions on the detector, and measured their centroids $(x_o, y_o)$ and $(x_e, y_e)$ respectively. For each pair, we computed the separation ($\sqrt{(x_o - x_e)^2 + (y_o-y_e)^2}$) and orientation ($\rm{arctan2}(y_o-y_e, x_o - x_e)$). We examined the separation and relative orientation of eo-pairs as a function of radial distance from the center of the CCD in Fig~\ref{Fig:B1} and Fig.~\ref{Fig:B2}, respectively. The mean separation is found to be $\sim 30.59$ pixels and is indicated by a red dashed line in Fig~\ref{Fig:B1}. 
The measured separations show no systematic dependence on radial distance and remain confined within the $\pm 1$ pixel (shown by the shaded region) of the mean value across the full field. Similarly, Fig.~\ref{Fig:B2}, representing the deviation of the orientation of eo-pairs from the mean value, exhibits no trend with the radial distance from the center. This confirms that the assumption of a constant separation is valid to first order over the entire CCD. To further improve sub-pixel accuracy, we include an optional recentering step in the photometric pipeline as discussed in Section~\ref{sec:3.2}.

\onecolumn
\section{Standard stars observations}\label{Ap:D}
\small
\begin{longtable}{llrrrrrrrrlc}
\caption{Observed and reference polarimetric parameters of standard stars.
The last column lists the literature source of the reference values.}
\label{tab:pol_appendix}\\

\toprule
Date & Object & $P_{\rm{pipe}}$ $\pm$ $\sigma_{P_{\rm{pipe}}}$ & $\theta_{\rm{pipe}}$ $\pm$
$\sigma_{\theta_{\rm{pipe}}}$ & 
$P_{\rm{lit}}$ $\pm$ $\sigma_{P_{\rm{lit}}}$ &
$\theta_{\rm{lit}}$ $\pm$ $\sigma_{\theta_{\rm{lit}}}$ & $\psi_{inst}$ & Reference \\
\midrule
\endfirsthead

\caption[]{Table~\ref{tab:pol_appendix} (continued)}\\
\toprule
Date & Object & $P_{\rm{pipe}}$ $\pm$ $\sigma_{P_{\rm{pipe}}}$ & $\theta_{\rm{pipe}}$ $\pm$
$\sigma_{\theta_{\rm{pipe}}}$ & 
$P_{\rm{lit}}$ $\pm$ $\sigma_{P_{\rm{lit}}}$ &
$\theta_{\rm{lit}}$ $\pm$ $\sigma_{\theta_{\rm{lit}}}$ & $\Delta\theta$ & Reference \\
\midrule
\endhead

\midrule
\multicolumn{12}{r}{\textit{Continued on next page}} \\
\bottomrule

\endfoot

\bottomrule

\midrule
\multicolumn{8}{p{\linewidth}}{\footnotesize
\textit{Note.}
References for Sc92, B23 W92, Sk18, and T90 are \citet{Schmidt1992}, \cite{Blinov2023},  \cite{Whittet1992}, \cite{Skalidis2018}, \cite{Turnshek1990} respectively.
}
\\
\endlastfoot
  26Nov2017 & BD+59389 & 6.41 $\pm$ 0.17 & -75.33 $\pm$ 0.74 & 6.43 $\pm$ 0.022 & 98.14 $\pm$ 0.1 & 173.47 &Sc92\\
  26Nov2017 & L111\_1965 & 1.2 $\pm$ 0.13 & 75.88 $\pm$ 3.09 & 1.216 $\pm$ 0.028 & 74.2  $\pm$ 0.7 & 178.32 &B23\\
  26Nov2017 & L111\_1969 & 1.14 $\pm$ 0.14 & 78.95 $\pm$ 3.44 & 1.229 $\pm$ 0.028 & 75.9 $\pm$ 0.6 & 176.95 &B23\\
  26Nov2017 & HD344776 & 6.57 $\pm$ 0.19 & 31.73 $\pm$ 0.74 & 6.101 $\pm$ 0.034 & 25.3 $\pm$ 0.2 & 173.57 &B23\\
  26Nov2017 & B0211p105137 & 0.69 $\pm$ 0.23 & 2.61 $\pm$ 9.48 & 0.761 $\pm$ 0.04 & -6.1 $\pm$ 1.5 &171.29 &B23\\
  26Nov2017 & B0211p105118 & 0.79 $\pm$ 0.2 & 14.02 $\pm$ 7.2 & 0.857 $\pm$ 0.044 &  1.4 $\pm$ 1.5 & 167.38 &B23\\
  27Nov2017 & BD+59389 & 6.48 $\pm$ 0.16 & -76.24 $\pm$ 0.9 & 6.43 $\pm$ 0.022 & 98.14 $\pm$ 0.1& 174.38 &Sc92\\
  27Nov2017 & L111\_1965 & 1.45 $\pm$ 0.19 & 86.48 $\pm$ 3.88 & 1.216 $\pm$ 0.028 & 74.2 $\pm$ 0.7 & 167.72 &B23\\
  27Nov2017 & L111\_1969 & 1.35 $\pm$ 0.14 & 80.52 $\pm$ 3.15 & 1.229 $\pm$ 0.028 & 75.9$\pm$ 0.6 & 175.38&B23\\
  27Nov2017 & HD344776 & 6.34 $\pm$ 0.19 & 32.26 $\pm$ 0.86 & 6.101 $\pm$ 0.034 & 25.3 $\pm$ 0.2 & 173.04 &B23\\
  27Nov2017 & B0211p105137 & 0.69 $\pm$ 0.25 & 17.28 $\pm$ 10.3 & 0.761 $\pm$ 0.04 & -6.1 $\pm$ 1.5 & 156.62 &B23\\
  27Nov2017 & B0211p105118 & 0.62 $\pm$ 0.21 & 24.17 $\pm$ 9.93 & 0.857 $\pm$ 0.044 &  1.4 $\pm$ 1.5 & 157.23 &B23\\
  27Nov2017 & B2015\_0137\_102 & 1.0 $\pm$ 0.89 & 89.66 $\pm$ 24.74 & 1.097 $\pm$ 0.029 & -82.0 $\pm$ 0.7 &8.34 &B23\\
  27Nov2017 & L\_95\_275 & 1.01 $\pm$ 0.19 & 58.65 $\pm$ 5.42 & 0.752 $\pm$ 0.044 & 46.2 $\pm$ 1.7 &167.55 &B23\\
  27Nov2017 & L\_95\_276 & 0.21 $\pm$ 0.31 & 36.33 $\pm$ 41.12 & 0.204 $\pm$ 0.051 & 44.8 $\pm$ 7.5 &8.47 & B23\\
  07Mar2018 & BD+312505 & 0.21 $\pm$ 0.13 & 79.77 $\pm$ 18.42 & 0.179 $\pm$ 0.033 & 52.6 $\pm$ 5.4 &152.83 &B23\\
  07Mar2018 & CMaR124 & 3.42 $\pm$ 0.35 & -86.31 $\pm$ 2.94 & 3.18 $\pm$ 0.09 & 86.0 $\pm$ 1 & 172.31&W92\\
  07Mar2018 & L\_97\_345 & 1.24 $\pm$ 0.17 & 5.85 $\pm$ 4.58 & 1.411 $\pm$ 0.064 & -0.5 $\pm$ 1.3 &173.65 &B23\\
  07Mar2018 & L\_97\_351 & 0.95 $\pm$ 0.12 & -35.57 $\pm$ 3.61 & 1.013 $\pm$ 0.094 & -35.1 $\pm$ 2.7 & 0.47&B23\\
  07Mar2018 & L\_98\_685 & 0.29 $\pm$ 0.25 & 89.3 $\pm$ 24.44 & 0.201 $\pm$ 0.114 &-9.5$\pm$ 20.6 & 81.2&B23\\
  07Mar2018 & L\_98\_653 & 0.42 $\pm$ 0.12 & -84.54 $\pm$ 8.02 & 0.439 $\pm$ 0.112 &  75.4 $\pm$ 7.6 & 159.94&B23\\  
  08Mar2018 & BD+312505 & 0.22 $\pm$ 0.18 & 68.7 $\pm$ 23.79 & 0.179 $\pm$ 0.033 & 52.6 $\pm$ 5.4 &163.9 &B23\\
  08Mar2018 & CMaR124 & 3.57 $\pm$ 0.37 & -82.05 $\pm$ 2.97 & 3.18 $\pm$ 0.09 & 86.0 $\pm$ 1 &168.05 &W92\\
  08Mar2018 & BD+332642 & 0.36 $\pm$ 0.23 & -76.27 $\pm$ 18.13 & 0.2 $\pm$ 0.15 & 78$\pm$ 20 & 154.27&Sk18\\
  08Mar2018 & L\_97\_345 & 1.19 $\pm$ 0.22 & 7.74 $\pm$ 5.39 & 1.411 $\pm$ 0.064 & -0.5 $\pm$ 1.3 &171.76 &B23\\
  08Mar2018 & L\_97\_351 & 0.96 $\pm$ 0.17 & -35.04 $\pm$ 5.03 & 1.013 $\pm$ 0.094 & -35.1 $\pm$ 2.7 & 179.94&B23\\
  08Mar2018 & L\_98\_685 & 0.30 $\pm$ 0.16 & 85.20 $\pm$ 16.01 & 0.201 $\pm$ 0.114 &-9.5$\pm$ 20.6 & 85.3&B23\\
  08Mar2018 & L\_98\_653 & 0.41 $\pm$ 0.11 & 87.54 $\pm$ 7.86 & 0.439 $\pm$ 0.112 &  75.4 $\pm$ 7.6 & 167.86&B23\\ 
  05May2022 & HD155197 & 4.36 $\pm$ 0.16 & 23.56 $\pm$ 1.04 & 4.274 $\pm$ 0.027 & 102.88 $\pm$ 0.18 & 79.32&Sc92\\
  05May2022 & HD154445 & 3.68 $\pm$ 0.09 & 11.76 $\pm$ 0.77 & 3.683 $\pm$ 0.072 & 88.91 $\pm$ 0.56 & 77.15&Sc92\\
  07May2022 & HD155197 & 4.35 $\pm$ 0.16 & 24.76 $\pm$ 1.01 & 4.274 $\pm$ 0.027 & 102.88 $\pm$ 0.18 & 78.12&Sc92\\
  07May2022 & Hiltner960 & 5.58 $\pm$ 0.19 & -22.7 $\pm$ 0.98 & 5.21 $\pm$ 0.029 & 54.54 $\pm$ 0.16 & 77.24&Sc92\\
  19Oct2022 & Hiltner960 & 5.53 $\pm$ 0.11 & -25.31 $\pm$ 0.59 & 5.21 $\pm$ 0.029 & 54.54 $\pm$ 0.16 & 79.85&Sc92\\
  19Oct2022 & BD+64106 & 5.36 $\pm$ 0.10 & 18.20 $\pm$ 0.55 & 5.15 $\pm$ 0.098 & 96.74 $\pm$ 0.54 & 78.54&Sc92\\
  19Oct2022 & HD19820 & 4.24 $\pm$ 0.04 & 36.67 $\pm$ 0.27 & 4.562 $\pm$ 0.025 & 114.46 $\pm$ 0.16 & 77.79&Sc92\\
  21Oct2022 & BD+64106 & 5.43 $\pm$ 0.14 & 17.54 $\pm$ 0.72 & 5.15 $\pm$ 0.098 & 96.74 $\pm$ 0.54 &79.20 &Sc92\\
  21Oct2022 & HD19820 & 4.42 $\pm$ 0.07 & 35.25 $\pm$ 0.47 & 4.562 $\pm$ 0.025 & 114.46 $\pm$ 0.16 & 79.21&Sc92\\
  21Oct2022 & HD7927 & 3.19 $\pm$ 0.06 & 11.15 $\pm$ 0.52 & 3.026 $\pm$ 0.037 & 90.84 $\pm$ 0.35 & 79.69&Sc92\\
  21Oct2022 & HD25443 & 4.71 $\pm$ 0.07 & 56.64 $\pm$ 0.42 & 4.734 $\pm$ 0.045 & 133.65 $\pm$ 0.28 & 77.01&Sc92\\
  21Oct2022 & HD236633 & 5.56 $\pm$ 0.08 & 12.64 $\pm$ 0.43 & 5.376 $\pm$ 0.028 & 93.04 $\pm$ 0.15 & 80.4&Sc92\\
  22Oct2022 & BD+59389 & 6.62 $\pm$ 0.06 & 19.27 $\pm$ 0.26 & 6.43 $\pm$ 0.022 & 98.14 $\pm$ 0.1 & 78.87&Sc92\\
  22Oct2022 & BD+59389 & 6.62 $\pm$ 0.06 & 19.27 $\pm$ 0.26 & 6.43 $\pm$ 0.022 & 98.14 $\pm$ 0.1 & 78.87&Sc92\\
  22Oct2022 & HD19820 & 4.72 $\pm$ 0.05 & 35.75 $\pm$ 0.33 & 4.562 $\pm$ 0.025 & 114.46 $\pm$ 0.16 & 78.71&Sc92\\
  23Oct2022 & BD+59389 & 6.59 $\pm$ 0.08 & 19.0 $\pm$ 0.34 & 6.43 $\pm$ 0.022 & 98.14 $\pm$ 0.1 & 79.14&Sc92\\
  23Oct2022 & BD+59389 & 6.59 $\pm$ 0.08 & 19.0 $\pm$ 0.34 & 6.43 $\pm$ 0.022 & 98.14 $\pm$ 0.1 & 79.14&Sc92\\
  23Oct2022 & BD+64106 & 5.52 $\pm$ 0.15 & 19.11 $\pm$ 0.75 & 5.15 $\pm$ 0.098 & 96.74 $\pm$ 0.54 & 77.63&Sc92\\
  23Oct2022 & HD19820 & 4.56 $\pm$ 0.06 & 36.28 $\pm$ 0.49 & 4.562 $\pm$ 0.025 & 114.46 $\pm$ 0.16 & 78.18&Sc92\\
  23Oct2022 & HD25443 & 4.74 $\pm$ 0.06 & 56.85 $\pm$ 0.42 & 4.734 $\pm$ 0.045 & 133.65 $\pm$ 0.28 & 76.8&Sc92\\
  02Dec2023 & BD+64106 & 5.25 $\pm$ 0.3 & 17.26 $\pm$ 1.65 & 5.15 $\pm$ 0.098 & 96.74 $\pm$ 0.54 & 79.48&Sc92\\
  02Dec2023 & HD19820 & 4.39 $\pm$ 0.13 & 34.79 $\pm$ 0.85 & 4.562 $\pm$ 0.025 & 114.46 $\pm$ 0.16 & 79.67&Sc92\\
  03Dec2023 & HD19820 & 4.39 $\pm$ 0.13 & 34.81 $\pm$ 0.83 & 4.562 $\pm$ 0.025 & 114.46 $\pm$ 0.16 & 79.65&Sc92\\
  11Feb2024 & HD251204 & 4.84 $\pm$ 0.47 & 74.29 $\pm$ 2.78 & 4.04 $\pm$ 0.066 &147   $\pm$ - & 72.71&T90\\
  29Oct2024 & HD204827 & 5.16 $\pm$ 0.08 & -23.06 $\pm$ 0.44 & 4.893 $\pm$ 0.029 & 59.1 $\pm$ 0.17 & 82.16&Sc92\\
  29Oct2024 & VICyg12 & 8.19 $\pm$ 0.28 & 33.62 $\pm$ 0.98 & 7.893 $\pm$ 0.037 & 116.23 $\pm$ 0.14 &82.61 &Sc92\\
  30Nov2024 & HD19820 & 4.45 $\pm$ 0.08 & 31.82 $\pm$ 0.49 & 4.562 $\pm$ 0.025 & 114.46 $\pm$ 0.16 & 82.64&Sc92\\
  02Dec2024 & BD+59389 & 6.31 $\pm$ 0.36 & 14.93 $\pm$ 1.63 & 6.43 $\pm$ 0.022 & 98.14 $\pm$ 0.1 &83.21 &Sc92\\
  04Dec2024 & HD25443 & 4.85 $\pm$ 0.44 & 51.57 $\pm$ 2.59 & 4.734 $\pm$ 0.045 & 133.65 $\pm$ 0.28 & 82.08&Sc92\\
  05Dec2024 & HD25443 & 5.01 $\pm$ 0.43 & 50.54 $\pm$ 2.45 & 4.734 $\pm$ 0.045 & 133.65 $\pm$ 0.28 &83.11 &Sc92\\
  01Mar2025 & HD251204 & 5.02 $\pm$ 0.23 & 71.72 $\pm$ 1.37 & 4.04 $\pm$ 0.066 & 147 $\pm$ - & 75.28&T90\\
  02Mar2025 & HD154445 & 3.5 $\pm$ 0.19 & 8.61 $\pm$ 1.55 & 3.683 $\pm$ 0.072 & 88.91 $\pm$ 0.56 &80.3 &Sc92\\
  05Mar2025 & HD154445 & 3.55 $\pm$ 0.23 & 7.88 $\pm$ 1.86 & 3.683 $\pm$ 0.072 & 88.91 $\pm$ 0.56 & 81.03&Sc92\\
  14Nov2025 & BD+64106 & 5.24 $\pm$ 0.17 & 5.7 $\pm$ 0.95 & 5.15 $\pm$ 0.098 & 96.74 $\pm$ 0.54 & 91.04&Sc92\\
  14Nov2025 & HD236633 & 5.26 $\pm$ 0.13 & 0.32 $\pm$ 0.73 & 5.376 $\pm$ 0.028 & 93.04 $\pm$ 0.15 & 92.72&Sc92\\
  15Nov2025 & BD+64106 & 5.32 $\pm$ 0.2 & 6.11 $\pm$ 1.1 & 5.15 $\pm$ 0.098 & 96.74 $\pm$ 0.54 &90.63 & Sc92\\
  15Nov2025 & HD236633 & 5.34 $\pm$ 0.15 & 0.37 $\pm$ 0.78 & 5.376 $\pm$ 0.028 & 93.04 $\pm$ 0.15 & 92.67&Sc92\\
  15Nov2025 & HD204827 & 5.1 $\pm$ 0.09 & -33.64 $\pm$ 0.5 & 4.893 $\pm$ 0.029 & 59.1 $\pm$ 0.17 &92.74 &Sc92\\
  15Nov2025 & HD251204 & 4.98 $\pm$ 0.19 & 61.87 $\pm$ 1.10 & 4.04 $\pm$ 0.066 & 147 $\pm$ -  & 85.13&T90\\
  16Nov2025 & BD+64106 & 5.33 $\pm$ 0.21 & 5.88 $\pm$ 1.13 & 5.15 $\pm$ 0.098 & 96.74 $\pm$ 0.54 & 90.86&Sc92\\
  16Nov2025 & HD19820 & 4.47 $\pm$ 0.05 & 23.55 $\pm$ 0.34 & 4.562 $\pm$ 0.025 & 114.46 $\pm$ 0.16 & 90.91&Sc92\\
  16Nov2025 & HD7927 & 3.06 $\pm$ 0.04 & -2.0 $\pm$ 0.34 & 3.026 $\pm$ 0.037 & 90.84 $\pm$ 0.35 & 92.84&Sc92\\
  16Nov2025 & HD236633 & 5.24 $\pm$ 0.11 & 0.46 $\pm$ 0.63 & 5.376 $\pm$ 0.028 & 93.04 $\pm$ 0.15 & 92.58&Sc92\\
  16Nov2025 & HD204827 & 5.27 $\pm$ 0.08 & -33.44 $\pm$ 0.41 & 4.893 $\pm$ 0.029 & 59.1 $\pm$ 0.17 & 92.54&Sc92\\
  17Nov2025 & BD+64106 & 5.32 $\pm$ 0.23 & 5.28 $\pm$ 1.23 & 5.15 $\pm$ 0.098 & 96.74 $\pm$ 0.54 & 91.46&Sc92\\
  17Nov2025 & HD236633 & 5.33 $\pm$ 0.15 & 0.87 $\pm$ 0.83 & 5.376 $\pm$ 0.028 & 93.04 $\pm$ 0.15 &92.17 &Sc92\\
  18Nov2025 & HD7927 & 3.09 $\pm$ 0.05 & -1.63 $\pm$ 0.47 & 3.026 $\pm$ 0.037 & 90.84 $\pm$ 0.35 &92.47 &Sc92\\
  27Nov2025 & HD236633 & 5.36 $\pm$ 0.4 & 3.02 $\pm$ 2.16 & 5.376 $\pm$ 0.028 & 93.04 $\pm$ 0.15 &90.02 &Sc92\\

\end{longtable}

\end{document}